\documentclass[a4paper,10pt,twoside]{cpc-hepnp}
\usepackage{multicol}
\usepackage{graphicx}
\usepackage{booktabs}
\usepackage{amssymb,bm,mathrsfs,bbm,amscd}
\usepackage[tbtags]{amsmath}
\usepackage{lastpage}

\usepackage{CJKutf8}
\usepackage[overlap, CJK]{ruby}
\newenvironment{SChinese}{%
\CJKfamily{gbsn}%
\CJKtilde
\CJKnospace}{}
\newcommand{\cntxt}[1]{\begin{CJK}{UTF8}{}\begin{SChinese}#1\end{SChinese}\end{CJK}}

\begin{document}

\fancyhead[c]{\small Chinese Physics C~~~Vol. xx, No. x (201x) xxxxxx}
\fancyfoot[C]{\small 010201-\thepage}

\footnotetext[0]{Received 31 June 2015}

\title{Electron Bunch Train Excited Higher-Order Modes in a Superconducting RF Cavity\thanks{Supported by National Natural Science Foundation of China (11275014)}}

\author{
      Yong-Feng Gao (\cntxt{高永凤})
\quad Sen-Lin Huang (\cntxt{黄森林})$^{1)}$\email{huangsl@pku.edu.cn}
\quad Fang Wang (\cntxt{王芳})
\quad Li-Wen Feng (\cntxt{冯立文})
\quad De-Hao Zhuang (\cntxt{庄德浩})\\
\quad Lin Lin (\cntxt{林林})
\quad Feng Zhu (\cntxt{朱凤})
\quad Jian-Kui Hao (\cntxt{郝建奎})
\quad Sheng-Wen Quan (\cntxt{全胜文})
\quad Ke-Xin Liu (\cntxt{刘克新})
}
\maketitle

\address{%
Institute of Heavy Ion Physics and State Key Laboratory of Nuclear Physics and Technology, School of Physics, \\
Peking University, Beijing 100871, China
}

\begin{abstract}
Higher-order mode (HOM) based intra-cavity beam diagnostics has been proved effectively and conveniently in superconducting radio-frequency (SRF) accelerators. Our recent research shows that the beam harmonics in the bunch train excited HOM spectrum, which have much higher signal-to-noise ratio than the intrinsic HOM peaks, may also be useful for beam diagnostics. 
In this paper, we will present our study on bunch train excited HOMs, including the theoretic model and recent experiments carried out based on the DC-SRF photoinjector and SRF linac at Peking University.
\end{abstract}



\footnotetext[0]{\hspace*{-3mm}\raisebox{0.3ex}{$\scriptstyle\copyright$}2013
Chinese Physical Society and the Institute of High Energy Physics
of the Chinese Academy of Sciences and the Institute
of Modern Physics of the Chinese Academy of Sciences and IOP Publishing Ltd}%

\begin{multicols}{2}

\section{Introduction}

Higher-order modes (HOMs) will be excited when electron beam passes through the radio-frequency (rf) cavity, which is one of the main causes of emittance dilution~\cite{AVostrikov2015} and beam current limit in a superconducting rf (SRF) cavity~\cite{RCalaga2003,SChen2013,SChen2015}. For TESLA cavity, two coaxial type couplers are therefore installed on both ends of the cavity with a specific orientation to extract HOMs and transfer the power to the load~\cite{JSekutowicz1993}.

Meanwhile, HOM signal contains beam information, such as the beam position in the rf cavity, the beam arrival time, and etc~\cite{JFrisch2006}. It can be directly used for beam diagnostics without additional vacuum instruments. Such HOM based beam diagnostics is convenient and cheap, and more importantly, it can provide local beam information in cryomodule. Many laboratories have conducted studies on HOM based beam diagnostics. At JAEA, HOM signal excited by macro pulse was detected to obtain beam position information~\cite{MSawamura2007}. At DESY, dipole mode signal excited by a single bunch was applied for SRF cavity alignment and beam position measurements~\cite{PZhang2012,LShi2015,SMolloy2007,NBaboi2011}.

In the spectrum of HOM signal excited by multiple bunches, two kinds of peaks exist: intrinsic HOM peaks and beam harmonics~\cite{SIKurokawa1999,SKim2002}. This can be observed clearly when the bunch repetition rate is high enough so that the beam harmonics can be identified by equipments. Under proper conditions, the beam harmonics are much stronger than the intrinsic HOM peaks, therefore they may provide a useful means to obtain beam information with high signal-to-noise ratio. 
In this paper, we will first present our theoretic model for bunch train excited HOMs and then the measurement results on the DC-SRF photoinjector and SRF linac. We will also report our recent tests of beam diagnostics on the DC-SRF photoinjector.  

\section{Electron Bunch Train Excited HOMs}

The HOM voltage excited by a relativistic electron beam in an SRF cavity can be calculated as 
\begin{equation}
V_{b}(t) = \sum_m\int_{-\infty}^{t}V_m(t-\tau)\rho_{b}(\tau)d\tau,
\end{equation}
where the summation is over all HOMs in the SRF cavity, $V_m$ is the single electron induced voltage signal of the $m^{th}$ HOM ($m = 1, 2, 3, ...$), $\rho_{b} = I_{b}/e$, $I_{b}$  is the current signal of electron beam and $e$ is the charge of an electron. Changing variable from $\tau$ to $\tau' = t-\tau$, we have 
\begin{equation}
V_{b}(t) = \sum_m\int_{0}^{\infty}V_m(\tau')\rho_{b}(t-\tau')d\tau'.
\end{equation}

The HOM magnitude spectrum can be calculated as
\begin{equation}
\mathbb{V}_{b}(f) = \lvert\hat{\rho}_{b}(f)\rvert\cdot\Bigl|\sum_m\int_{0}^{\infty}V_m(\tau')e^{-2\pi i f\tau'}d\tau'\Bigr|,
\label{eq:HOM:spec:0}
\end{equation}
where 
\begin{equation}
\hat{\rho}_{b} = \int_{-\infty}^{\infty}\rho_{b}(t)e^{-2\pi i f t}dt.
\end{equation}
Considering the HOM wake field excited by a single electron exists only when $\tau'>0$, the lower limit of integration in Eq. ($\ref{eq:HOM:spec:0}$) can be extended to $-\infty$ and as a result, 
\begin{equation}
\mathbb{V}_{b}(f) = \lvert\hat{\rho}_{b}(f)\rvert\cdot\Bigl|\sum_m\hat{V}_m(f)\Bigr|,
\end{equation}
in which 
\begin{equation}
\hat{V}_m(f) = \int_{-\infty}^{\infty}V_m(\tau')e^{-2\pi i f\tau'}d\tau', 
\end{equation}
is the single electron excited HOM spectrum. Referring to, e.g.,~\cite{AWChao1993,SLAC-PUB-10792},  it can be calculated as 
\begin{equation}
\hat{V}_m(f) = e \alpha_m Z_m^{\parallel}(f),
\end{equation}
in which $\alpha_m$ is the mode excitation factors of the $m^{th}$ HOM, 
accounting for the coupling efficiency between the HOM field and electron energy. 
For dipole mode~\cite{HPadamsee1998}, e.g., $\alpha_m$ accounts for the coupling efficiency due to the transverse offset of electron beam trajectory in the SRF cavity. 
$Z_m^{\parallel}(f)$ in the equation is the longitudinal impedance of the $m^{th}$ HOM, 
which can often be modelled by an equivalent parallel LRC resonator circuit and given by~\cite{AWChao1993}
\begin{equation}
Z_m^{\parallel}(f)=\frac{(R/Q)_m}{1/Q_m+i (f_m/f-f/f_m)},
\end{equation}
where $Q_m$ and $f_m$ are the quality factor and eigenfrequency of the $m^{th}$ HOM, respectively. Therefore we have
\begin{equation}
\mathbb{V}_b(f) = \mathbb{I}_b(f)\cdot\Bigl|\sum_m\frac{\alpha_m (R/Q)_m}{1/Q_m+i (f_m/f-f/f_m)}\Bigr|,
\label{eq:HOM-spec-formula-complete}
\end{equation}
in which $\mathbb{I}_b(f) = \lvert\hat{I}_b(f)\rvert$, and
\begin{equation}
\hat{I}_{b}(f)=e\hat{\rho}_{b}(f)=\int_{-\infty}^{\infty}I_{b}(t)e^{-2\pi i f t}dt
\end{equation}
is the current spectrum of the electron beam.

Introducing ${\delta{f}}_m = f - f_m$, we have
\begin{equation}
\mathbb{V}_b(f) = \mathbb{I}_b(f)\cdot\Bigl|\sum_m\frac{\alpha_m (R/Q)_m}{1/Q_m-i \left(\frac{{\delta f}_m}{f_m+{\delta f}_m}+\frac{{\delta f}_m}{f_m}\right)}\Bigr|.
\end{equation}
The equation can be approximated as
\begin{equation}
\mathbb{V}_b(f) = \Bigl|\sum_m \alpha_m\mathbb{I}_b(f)\cdot\frac{f_m}{2\,{\delta{f}}_m} (R/Q)_m\Bigr|
\label{eq:HOM-spec-formula-approx}
\end{equation}
when $1\gg{\delta{f}}_m/f_m\gg 1/Q_m$.

Considering the electron beam comprising of a train of electron bunches with the same current profiles and equally separated by $T_b$ in time domain, the current signal can be represented using 
\begin{equation}
I_{b}(t)=\sum_{n=0}^{N-1}\int_{-\infty}^{\infty}I_{1}(\tau)\delta(t-nT_{b}-\tau)d\tau, 
\end{equation}
where $N$ is the number of bunches in the train, $I_1(t)$ represents the current profile of a single bunch, and $\delta(t)$ is the Dirac Delta function. 
It is easy to obtain the amplitude of current spectrum as
\begin{equation}
\mathbb{I}_{b}(f) = \Bigl|\frac{\sin(\pi N f T_{b})}{\sin(\pi f T_{b})}\Bigr|\cdot\mathbb{I}_{1}(f), 
\label{eq:bunch-train-current-spec}
\end{equation}
in which $\mathbb{I}_1(f) = \lvert\hat{I}_1(f)\rvert$, and
\begin{equation}
\hat{I}_{1}(f)=\int_{-\infty}^{\infty}I_{1}(t)e^{-2\pi i f t}dt
\end{equation}
is the Fourier transform of $I_1(t)$. 
In the case of electron bunch length $l_b \ll 1/f_n$, $\mathbb{I}_1(f)$ can be treated as a constant. 

As an example, we calculated a typical spectrum of HOM excited by a picosecond electron bunch train using Eq. ($\ref{eq:HOM-spec-formula-complete}$) and Eq. ($\ref{eq:bunch-train-current-spec}$). To clearly illustrate the characteristic of bunch train excited HOMs, only one HOM is considered in the calculation. 
The parameters are taken according to our experiments reported in the following: the HOM has an eigenfrequency of 1.7513 GHz and a quality factor of $5\times 10^5$, while the electron beam has a pulse repetition rate ($f_b = 1/T_b$) of 0.797 MHz and a macro-pulse length of 3 ms.

Fig.~$\ref{fig:electron-HOM-spec}$ - Fig.~$\ref{fig:bunch-train-HOM-spec}$ show the HOM spectrum excited by a single electron, the current spectrum of the bunch train, and the bunch train excited HOM spectrum, respectively.
It can be seen from Fig.~$\ref{fig:bunch-train-HOM-spec}$ that, besides the intrinsic HOM peak (referred to as ``intrinsic HOM peak'' herein), many spectral peaks (lines) at beam harmonic frequencies (referred to as ``beam harmonics'' herein) exist. Moreover, the beam harmonic closest to the intrinsic HOM peak has a much larger amplitude compared to the intrinsic HOM peak.

The HOM spectrum excited by a single electron shown in Fig.~$\ref{fig:electron-HOM-spec}$ is essentially the SRF cavity's spectral response function to an rf source, while the electron beam plays the role of an rf source with the spectrum shown in Fig.~$\ref{fig:curr-spec}$.
Although the SRF cavity's spectral response function has a very sharp spike at the HOM's eigenfrequency and the ratio between the amplitudes at eigenfrequency and beam harmonics is significant, the rf source, however, has an even higher ratio between the amplitudes at beam harmonics and eigenfrequency. Therefore, many beam harmonics higher than the intrinsic HOM peak can be observed. Obviously, the beam harmonic closest to the intrinsic HOM peak has the maximum amplitude, which can also be inferred from Eq.~(\ref{eq:HOM-spec-formula-approx}) directly. 

\begin{center}
\includegraphics[width=0.4\textwidth]{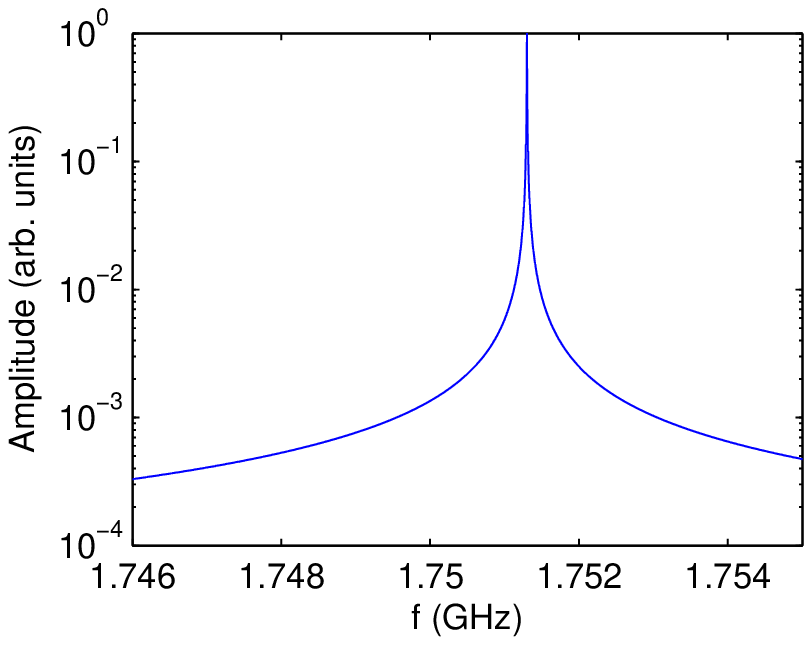}
\figcaption{\label{fig:electron-HOM-spec}  
Calculated HOM spectrum excited by a single electron. In the calculation, only one HOM is considered, which has an eigenfrequency of 1.7513 GHz and a quality factor of $5\times 10^5$.}
\end{center}

\begin{center}
\includegraphics[width=0.4\textwidth]{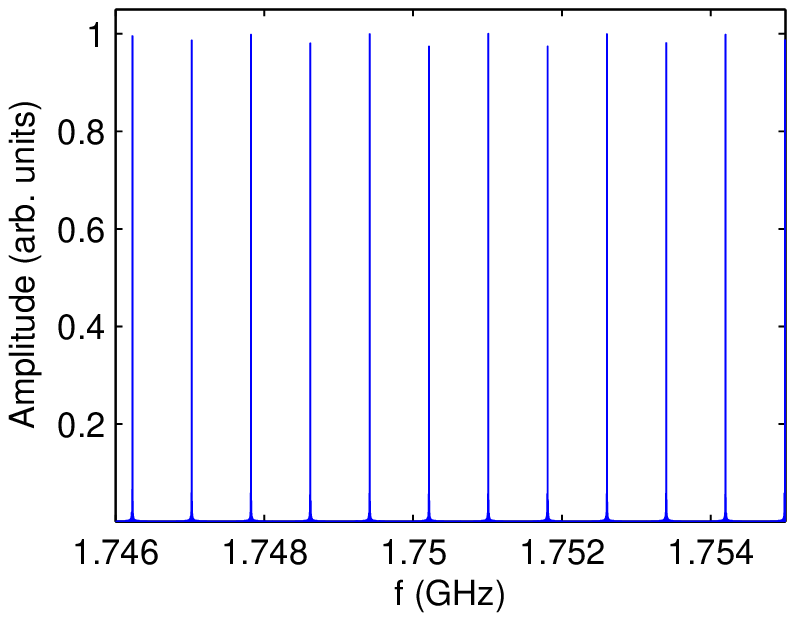}
\figcaption{\label{fig:curr-spec}  
Calculated current spectrum of a train of electron bunches. In the calculation, the electron beam has a pulse repetition rate of 0.797 MHz and a macro-pulse length of 3 ms.}
\end{center}

It is straightforward to infer that, in the case of only one HOM, the amplitude of beam harmonics has a linear dependency on the coupling efficiency $\alpha_m$ between the HOM field and electron energy. When the HOM is a dipole mode, the beam harmonic amplitude is inevitably proportional to the beam offset and therefore it can provide the beam position information, as that can be provided by the intrinsic HOM excited by a single electron bunch. Compared to the latter case, the beam harmonic has a much higher amplitude, which can provide a much higher signal-to-noise ratio, especially when the HOM quality factor $Q_m$ is small. 

\begin{center}
\includegraphics[width=0.4\textwidth]{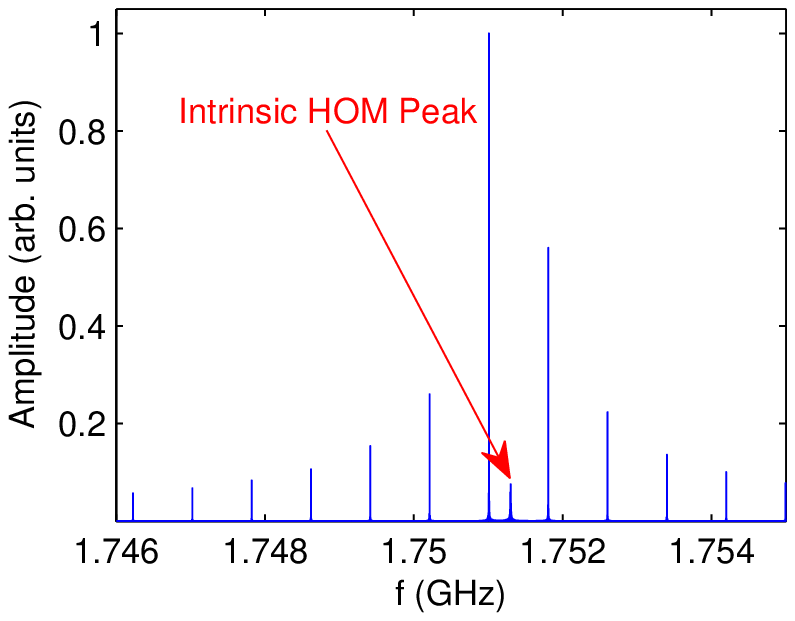}
\figcaption{\label{fig:bunch-train-HOM-spec}  
Calculated bunch train excited HOM spectrum. In the calculation, only one HOM is considered, which has an eigenfrequency of 1.7513 GHz and a quality factor of $5\times 10^5$. The electron beam has a pulse repetition rate of 0.797 MHz and a macro-pulse length of 3 ms.}
\end{center}

In practice, multiple HOMs could be excited in the SRF cavity. One can infer from Eq.~($\ref{eq:HOM-spec-formula-approx}$) and Eq.~($\ref{eq:bunch-train-current-spec}$) that, different HOMs would contribute to a same beam harmonic. However, the main contribution comes from the mode with higher $R/Q$ and with the eigenfrequency in proximity of the beam harmonics. With a properly chosen HOM, which has an eigenfrequency close to the beam harmonics and an $R/Q$ much higher than neighboring HOMs, the above discussion for only one HOM case is still valid. Under such a condition, it is obvious that the beam harmonics can be used for beam diagnostics.
 
 \section{HOM Experiments on the DC-SRF Photoinjector}
 
Preliminary experiments have been carried out based on the DC-SRF photoinjector at Peking University. 
The DC-SRF photoinjector comprises a pair of DC Pierce electrodes  and a 3.5-cell 1.3 GHz TESLA-type cavity~\cite{DCSRF2010}. It has been in operation since 2014 and can deliver 3 - 5 MeV, 6 - 50 pC electron bunches with the bunch length of approximately 5 ps and repetition rate tunable between 162.5 kHz and 81.25 MHz~\cite{DCSRF2015,DCSRF2016}. 
In the DC-SRF photoinjector, the electron bunches emitted from the photocathode are first accelerated by the DC Pierce structure to 50 - 90 keV. After a drift section of approximately 17 mm, the bunches enter the 3.5-cell SRF cavity. They are soon accelerated  to $\beta$ close to 1 in the first half cell. Therefore, the effect of $\beta$ variation on the electron beam excited HOM~\cite{KurennoySS1999} can be ignored in our experiments. 

The HOM experiments were performed when the DC-SRF photoinjector was operated in pulse mode with a macro pulse length of a few milliseconds. 
The electron beam excited HOM signal in the 3.5-cell SRF cavity was extracted from the pickup port of the cavity and then amplified by a three-stage power amplifier, which had a gain of 30 dB over the frequency range from 1.0 GHz to 2.4 GHz, before being delivered to the spectrum analyzer. 
Fig.~$\ref{fig:HOM-spec-meas}$ shows an example of the measured HOM spectrum, in which case the electron beam has a bunch repetition rate of 0.797 MHz and a macro pulse length of 3 ms.  In the measurement, the spectrum analyzer was set to sweep across the frequency range from 1.746 to 1.755 GHz, in which the $\textrm{TE}_{111}$ $3\pi/4$ mode, whose $R/Q$  is much higher than the neighboring HOMs in the SRF cavity, is expected. 

The result is consistent with that obtained in analytical calculation (see Fig.~$\ref{fig:bunch-train-HOM-spec}$ ). In Fig.~$\ref{fig:HOM-spec-meas}$, four discrete spectral lines equally spaced by 0.797 MHz can be clearly observed, corresponding to beam harmonics. The intrinsic HOM peak, however, is at noise level and cannot be observed.  From the figure one can infer that the eigenfrequency of the mode is between 1.7509 and 1.7516 GHz,corresponding to the frequencies of the two highest beam harmonics, respectively.

\begin{center}
\includegraphics[width=0.4\textwidth]{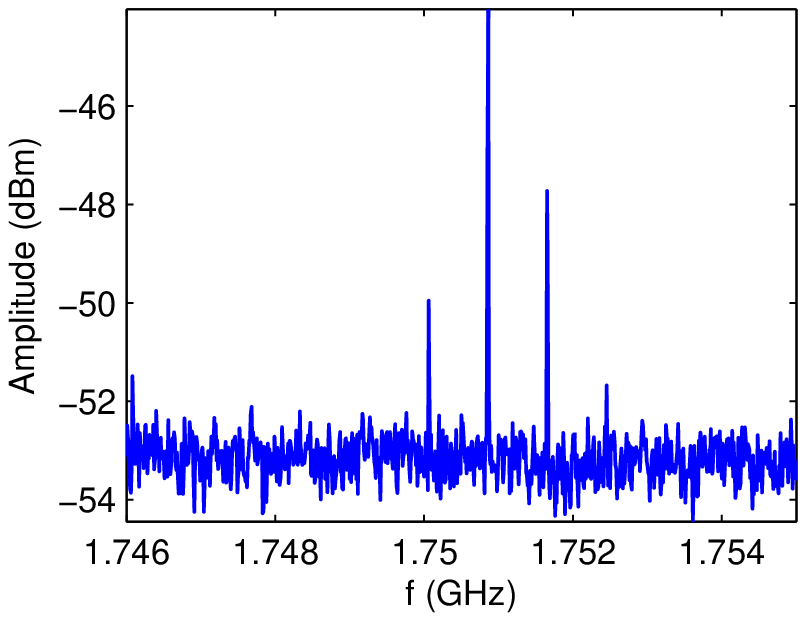}
\figcaption{\label{fig:HOM-spec-meas}  
Measured electron beam excited HOM spectrum in the 3.5-cell cavity of the DC-SRF photoinjector. In the measurement, the electron beam has a pulse repetition rate of 0.797 MHz and a macro-pulse length of 3 ms.}
\end{center}

Among the HOMs, dipole modes are related to beam position in the SRF cavity. Therefore preliminary investigation on the relationship between electron beam position and dipole modes was carried out. Three sets of tests were performed, in which $\textrm{TE}_{111}$ $3\pi/4$, $\textrm{TM}_{110}$ $\pi$, and $\textrm{TM}_{110}$ $\pi/4$ modes, which have the highest $R/Q$ among all dipole modes, are measured, respectively. The electron bunch repetition rate was set to be 0.797 MHz for $\textrm{TE}_{111}$ $3\pi/4$ mode measurement and 0.406 MHz for $\textrm{TM}_{110}$ $\pi$ mode and $\textrm{TM}_{110}$ $\pi/4$ mode measurements.

In all the three measurements, the beam harmonics with the largest amplitude were chosen for analysis, whose frequencies are 1.7509 GHz (for $\textrm{TE}_{111}$ $3\pi/4$ mode), 1.8013 GHz (for $\textrm{TM}_{110}$ $\pi$ mode), and 1.8875 GHz (for $\textrm{TM}_{110}$ $\pi/4$ mode), respectively. The  electron beam position in the SRF cavity was scanned through adjusting the position of drive laser~\cite{ZWang2016} spot on the photocathode. The whole scanning range of the electron beam position is estimated to be 8 mm.
The measured beam harmonic amplitudes, scaled by the electron beam current, are plotted as a function of the electron beam position in Fig.~$\ref{fig:HOM-pow-vs-pos}$. We need to point out that, although the drive laser position on the photocathode can be determined in experiments, the beam position in the SRF cavity, however, is difficult to be preciously calibrated at the moment. Therefore, the point numbers of scanning is used in Fig.~$\ref{fig:HOM-pow-vs-pos}$ to represent different beam positions.

\begin{center}
\includegraphics[width=0.4\textwidth]{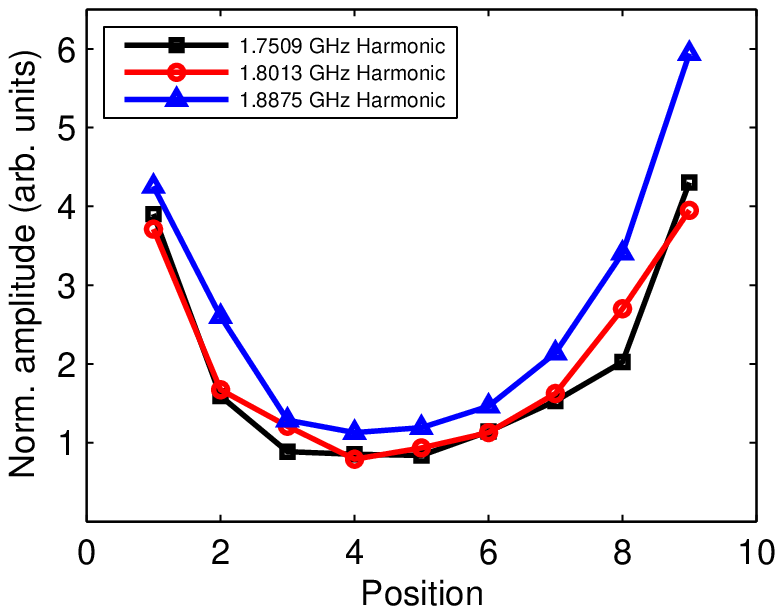}
\figcaption{\label{fig:HOM-pow-vs-pos}  
Measured electron beam excited HOM signal (beam harmonic) amplitude as a function of the electron beam position in the 3.5-cell SRF cavity of the DC-SRF photoinjector. 
Position number is used instead of the real position of the electron beam in the 3.5-cell cavity since the absolute position has not been calibrated.}
\end{center}

From Fig.~$\ref{fig:HOM-pow-vs-pos}$ one can find that, in all the three sets of tests, the beam harmonic amplitude has the minimum value  between position 4 and 5, which can be considered as the cavity center, and it grows when the electron beam deviates from the cavity center. This proves that the beam harmonics can be used for beam position measurement.

\section{HOM Signal from a 9-cell SRF Cavity}

Recently an SRF linac designed by Peking University was brought into operation to further accelerate the electron beam from the DC-SRF photoinjector. The linac~\cite{FZhu2015} consists of two 1.3-GHz 9-cell TESLA-type cavities. At either side of the 9-cell cavities two HOM couplers are mounted with an azimuth angle of $115^\circ$. The HOM signal in the 9-cell SRF cavity can be extracted from the HOM coupler.

\begin{center}
\includegraphics[width=0.4\textwidth]{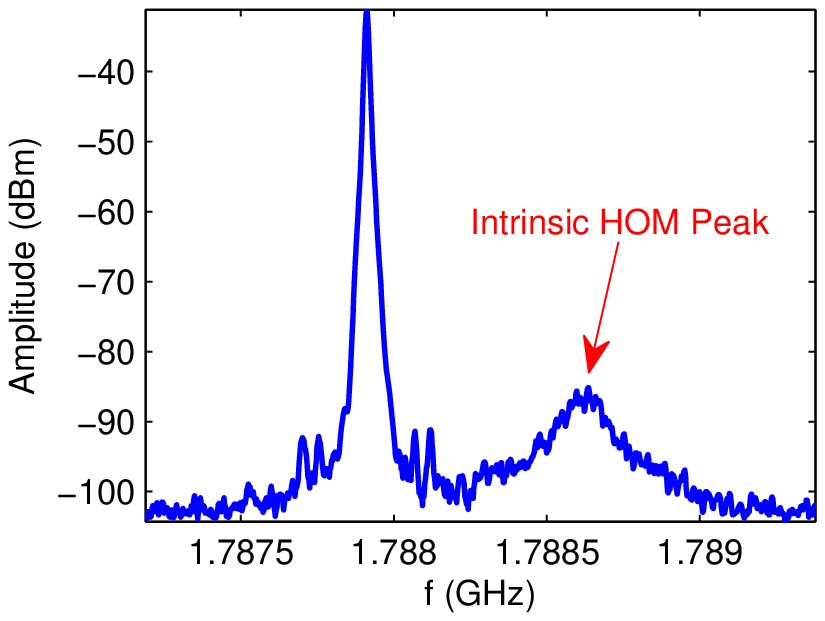}
\figcaption{\label{fig:HOM-spec-meas-2}  
Measured spectrum of electron beam excited HOM signal extracted from the HOM coupler. In the measurement, the electron beam has a pulse repetition rate of 27 MHz and a macropulse length of 3 ms.}
\end{center}

Spectrum measurements of the bunch train excited HOMs have been carried out based on the SRF linac. Similar to the above DC-SRF experiments, the linac was operated in pulse mode with a macro pulse length of a few milliseconds. Fig.~$\ref{fig:HOM-spec-meas-2}$ shows a measured spectrum, in which case the spectrum analyzer was set to sweep across the frequency range around the $\textrm{TE}_{111}$ $\pi$ mode.  As can be seen in the figure, both the beam harmonic and the intrinsic HOM peak can be observed, and the beam harmonic has a much higher (five orders of magnitude) amplitude and a much narrower band. 
During the experiment, we have also observed that the intrinsic HOM peak drifted randomly, which might be attributed to the cavity vibration, while the beam harmonic kept unchanged. This indicates that beam harmonics have the advantage in those applications requiring stable HOM spectral performance.

\section{Conclusion}

In this paper we have presented our study on bunch train excited HOMs in an SRF Cavity. Compared to the HOMs excited by single electron bunch, the bunch train excited HOMs have distinct spectral feature, which has been shown both theoretically and experimentally. 
Under proper conditions, beam harmonics are much stronger than the intrinsic HOM peaks. Therefore the beam harmonics could provide a useful means to obtain electron beam information with high signal-to-noise ratio.
Preliminary experiments demonstrate the usage of dipole mode beam harmonics in electron beam position diagnostics in the SRF cavity. Further research on HOM based beam diagnostics will be carried out.

\end{multicols}

\vspace{-1mm}
\centerline{\rule{80mm}{0.1pt}}
\vspace{2mm}

\begin{multicols}{2}

\end{multicols}

\clearpage
\end{document}